\documentclass[conference]{IEEEtran}
\usepackage[latin9]{inputenc}
\usepackage{amsmath}
\usepackage{graphicx,subfigure}
\usepackage{algorithm, algorithmic}
\usepackage{multirow}
 
\ifCLASSINFOpdf
\else
\fi
 
\usepackage{stfloats}

\begin{document}
\title{Enhancing JPEG Steganography using Iterative Adversarial Examples}
 
\author{\IEEEauthorblockN{Huaxiao Mo, Tingting Song,  Bolin Chen, Weiqi Luo*}
\IEEEauthorblockA{Guangdong Key Lab of Information Security \\Sun Yat-Sen University, Guangzhou, P.R. China \\
Email:  luoweiqi@mail.sysu.edu.com }
\and
\IEEEauthorblockN{Jiwu Huang}
\IEEEauthorblockA{Guangdong Key Lab of Intelligent Information Processing \\ 
Shenzhen University,  Shenzhen, P.R. China\\
Email:  jwhuang@szu.edu.cn}}
 \maketitle

\begin{figure}[b]
\vspace{-0.3cm}
\parbox{\hsize}{\em
WIFS`2019, December, 9-12, 2019, Delft, Netherlands.
XXX-X-XXXX-XXXX-X/XX/\$XX.00 \ \copyright 2019 IEEE.
}\end{figure}

\begin{abstract}
Convolutional Neural Networks (CNN)  based methods have significantly  improved the performance of image steganalysis compared with  conventional ones based on hand-crafted features.  However,  many existing literatures on computer vision have pointed out that those effective CNN-based methods  can be easily fooled by adversarial examples \cite{szegedy2013intriguing}.  In this paper, we propose a novel steganography framework based on adversarial example in an iterative manner.   The proposed framework first starts from an existing embedding cost,  such as  J-UNIWARD \cite{holub2014universal} in this work,  and then updates the cost iteratively based on  adversarial examples derived from  a series of steganalytic networks until achieving satisfactory results.  We carefully analyze two important factors that would affect the security performance of the proposed framewrork, i.e. the percentage of selected gradients with larger amplitude and the adversarial intensity to modify embedding cost.  The experimental results evaluated on three modern steganalytic models, including GFR, SCA-GFR and SRNet,  show that the proposed framework is very promising to enhance the security performances of  JPEG steganography.  
\end{abstract}

\IEEEpeerreviewmaketitle

\section{Introduction}
\label{sec:Introduction}
 
 Image steganography aims to hide secret message into digital image in an imperceptible manner, which has attracted much attention in the past decade. Up to now,  many steganography methods have been proposed. 
 
Most modern steganography methods are  constructed under the framework of distortion minimization,  such as  S-UNIWARD \cite{holub2014universal} and HILL \cite{HILL} in spatial domain,   J-UNIWARD \cite{holub2014universal} and  UERD \cite{UERD} in JPEG domain.  The most important issue in these methods is  to assign proper costs for all  embedding units (pixels / DCT coefficients). Then,  the common operation -  syndrome-trellis coding (STC)  \cite{STC}  -  is  used for subsequent data embedding.   However, most existing costs are usually empirical and  symmetrical (i.e. the cost for $+1$ and $-1$ for any embedding unit are exactly the same).   Taking J-UNIWARD for example,  it firstly uses three fixed wavelet directional filters to assess the texture complexity of cover image, and then defines the embedding costs of both $+1$ and $-1$ as the sum of relative changes of wavelet coefficients.  In this way,  embedding modifications will be located at those regions with  higher complexity that seem hard to model.  To enhance  existing steganography significantly,  some asymmetry costs based on data-driven methods  are worth studying.  

Recently, CNN-based steganalytic methods, such as  \cite{Covariance2019},  \cite{SRNet},  can achieve the state-of-the-art results on image steganalysis  compared with  conventional methods based on hand-crafted features.   However, many existing literatures in computer vision,  such as \cite{szegedy2013intriguing,goodfellow2015explaining},  pointed out  that  an adversarial example designed for a targeted CNN-based model  is often useful for fooling other models with different structures. Inspired by this,  several steganography methods  based on  adversarial examples  have been proposed.    In \cite{ma2018weakening}, Ma et al proposed a method  generating steganographic adversarial examples according to gradient map and embedding modification positions in spatial domain.  In \cite{Adversarial_IFS2019},  Tang et al proposed a JPEG stegnogarphy based on adversarial example.  In this method,   the DCT coefficients are divided  into two non-overalpping parts randomly, and then  data embedding is performed in two steps.  Firstly,  the method embeds secret message into the first part using  J-UNIWARD \cite{holub2014universal}.  Secondly,  it updates the embedding cost in the other part according to adversarial example, and then embeds the rest message into this part.   In \cite{zhang_adversarial2018},  Zhang et al constructed enhanced covers by adding adversarial noise to cover based on adversarial example technique.  Those enhanced covers have better characteristics for enhancing the security of  existing steganography in spatial domain.   Although the existing methods based on adversarial example can outperform the corresponding steganography,  there is a lots of room for further improvement based on our experiments. 

In this paper,  we propose a novel steganography framework based on adversarial example in an iterative manner.  We first set the initial embedding cost of input cover images using some existing steganography methods, such as 
such as J-UNIWARD \cite{holub2014universal} in this paper,  we  then update the  embedding cost  iteratively using adversarial attack to a  CNN based steganalyzer until achieving satisfactory results.   Experimental results show that the proposed  framework can significantly enhance the existing JPEG steganography evaluated on both deep learning based steganalyzers and the  conventional ones.   Please note that  although both our method  and the  method in \cite{Adversarial_IFS2019} aim to  enhance J-UNIWARD  based on  adversarial examples,   there are some important differences between them:  1) With iterative adversarial attacks,  we aim to update  embedding costs multiple times,  while the method \cite{Adversarial_IFS2019}  aims to  find a proper amount of adjust elements;   2)  We embed the whole secret message into JPEG directly, while the method  \cite{Adversarial_IFS2019}  processed it  in two batches;   3)  We just update those embedding costs with  larger gradient amplitudes  rather than  a random way in \cite{Adversarial_IFS2019};   4)  For updating embedding cost,  we use an additive way  instead of multiplicative one in \cite{Adversarial_IFS2019}.   

The rest of this paper is arranged as follows. Section \ref{sec:Proposed} describes the proposed  steganpgraphy framework. Section \ref{sec:Results} shows the   results and analysis.  Finally, the concluding remarks of this paper  would be given in Section \ref{sec:Conclusion}.    

\section{Proposed Steganography Framework}
\label{sec:Proposed}
The proposed steganography framework includes two stages, that is,  training a steganography model using iterative adversarial attacks and generating stego images based on the pre-trained steganography model.  Please note that images employed in the two stages  are different.

\begin{figure*}[t!]
    \centering
     \includegraphics[width=7in]{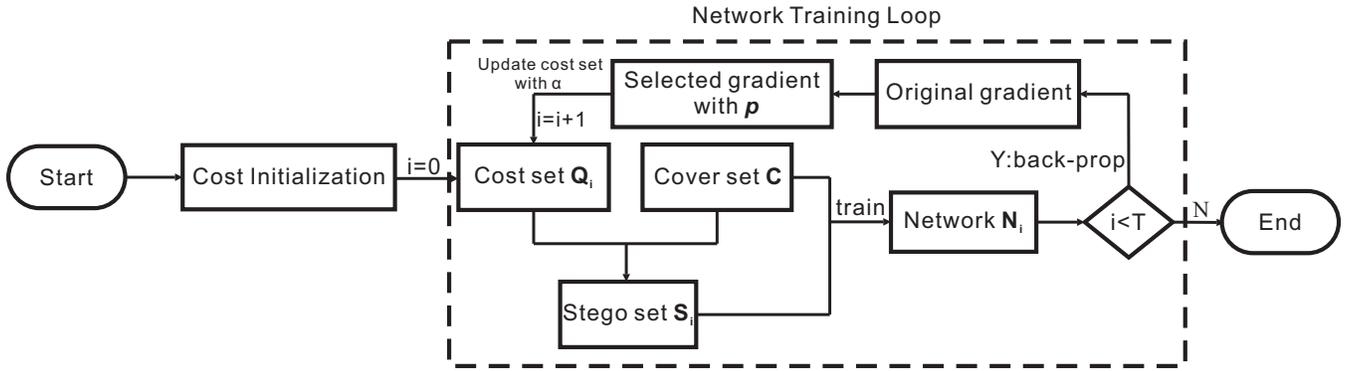}
    \caption{The training stage in the proposed steganography framework}
    \label{fig:Steganography_framework}
\end{figure*}

\begin{figure*}[t!]
    \centering
     \includegraphics[width=7in]{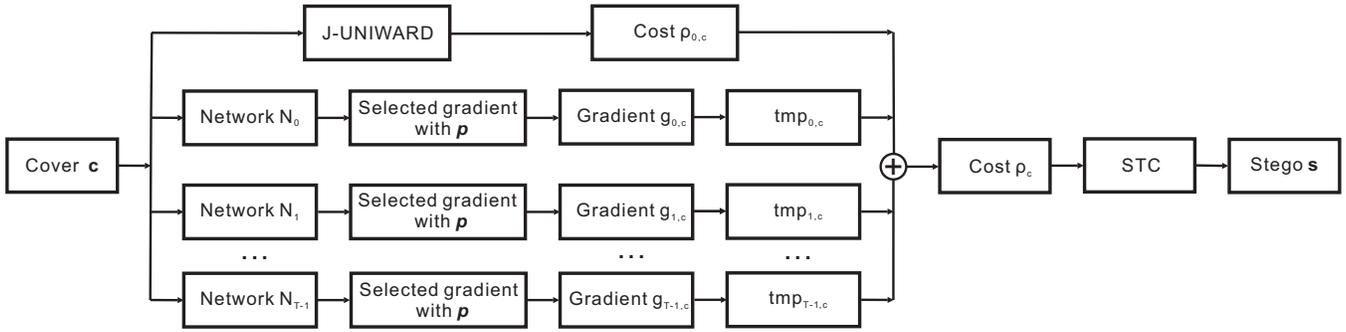}
    \caption{The process of generating image stego}
    \label{fig:generating_stego}
\end{figure*} 

\subsection{Training a Steganography Model}
\label{subsec:Training} 
As illustrated in Fig. \ref{fig:Steganography_framework},  the proposed framework firstly obtains the initial embedding cost set $Q_0$ of cover image set $C$  using some existing steganography method,  such as J-UNIWARD.  And then the cost set will be updated $T-1$  times in the proposed network training loop, where $T\geq2$.  For the $i^{th}$  iteration ($i=0,1 \ldots  T-1$),  we first obtain  image stego set $S_i$  according to the cover image set $C$ and their embedding cost set $Q_i$.  We then train a steganalytic network $N_i$ evaluated on the two image sets, i.e. $(C, S_i)$.  Based on the network $N_i$, we can obtain the corresponding gradients of the embedding units (i.e. DCT coefficients) for each cover image in $C$ by back-propagation.   For the $j^{th}$  cover in $C$,  denoted as $c_j$,  we first sort the absolute gradients of its DCT coefficients  in descending order,  and then select top $p$ gradients,  where $ 0<p\leq1$.   Finally,  we update the corresponding embedding costs with the selected gradients based on the following formulas.     

\begin{equation}
\label{eqa:1}
\rho_{i+1,j}^{+}(x,y) =
\begin{cases}
\rho_{i,j}^{+}(x,y) + 1 & \text{$g_{i,j}(x,y) < 0$} \\
\rho_{i,j}^{+}(x,y) + \alpha & \text{$g_{i,j}(x,y) > 0$}
\end{cases}
\end{equation}
\begin{equation}
\label{eqa:2}
\rho_{i+1,j}^{-}(x,y) =
\begin{cases}
\rho_{i,j}^{-}(x,y) + \alpha & \text{$g_{i,j}(x,y) < 0$} \\
\rho_{i,j}^{-}(x,y) + 1 & \text{$g_{i,j}(x,y) > 0$}
\end{cases}
\end{equation}

where $\rho^{+/-}_{i,j}(x,y)$ denotes the embedding cost for the $j^{th}$ cover in cover set $C$ in the $i^{th}$ iteration;  $(x,y)$ denotes the coordinate of embedding cost to be updated; superscript (i.e. $+$ or $-$) denotes the modification direction \footnote{Most existing steganography methods employ a symmetrical embedding scheme, which assumes that  the cost of modifying each embedding unit  by $+1$ and $-1$ are  the same, i.e. $\rho = \rho^{+} =\rho^{-} $.  However, the proposed steganography uses an asymmetry way, which can be observed from  the equations (1) and (2). }; the parameter $\alpha$ denotes the adversarial intensity for adjusting the cost, where $\alpha >1$ ;  $g_{i,j}(x,y)$ denotes the corresponding selected gradient  with the network  $N_i$.

\subsection{Generating Stegos Based on Pre-trained Model}
\label{subsec:Creating_Stegos}

The process of generating image stego is as illustrated in Fig. \ref{fig:generating_stego}.  For a given JPEG cover $c$,  we first obtain its embedding cost using J-UNIWARD,  denoted as $\rho_{0,c}$ . And then the cover is fed to the $T$ steganalytic networks obtained in  previous stage  (i.e. $N_i, i = 0, \ldots , T-1$)  to obtain  the corresponding  gradients $g_i$ with a parameter $p$,  and calculate $T$ temporary costs $tmp_{i,c}$  according to the following equation: 

\begin{equation}
\label{eqa:3}
tmp_{i,c}^{+}(x,y) =
\begin{cases}
1 & \text{$g_{i,c}(x,y) < 0$} \\
\alpha & \text{$g_{i,c}(x,y) > 0$}
\end{cases}
\end{equation}
\begin{equation}
\label{eqa:4}
tmp_{i,c}^{-}(x,y) =
\begin{cases}
\alpha & \text{$g_{i,c}(x,y) < 0$} \\
1 & \text{$g_{i,c}(x,y) > 0$}
\end{cases}
\end{equation}

Finally, the proposed embedding cost $\rho_c$ of  the input cover c is defined as follows:
\begin{equation}
\label{eqa:5}
\rho_{c}^{+} = \rho_{0,c} + \sum_{i=0}^{T-1} tmp_{i,c}^{+}; \rho_{c}^{-} = \rho_{0,c} + \sum_{i=0}^{T-1} tmp_{i,c}^{-}
\end{equation} 
  
 From the formulas (1)-(5), we observe that in each round of cost updating,  the embedding cost will become relatively larger ($\alpha>1$) when its modification ($+$ or $-$) and the corresponding gradient ($g>0$ or $g<0$) have the same direction.

\begin{figure*}[t!]
    \centering
     \includegraphics[width=6in]{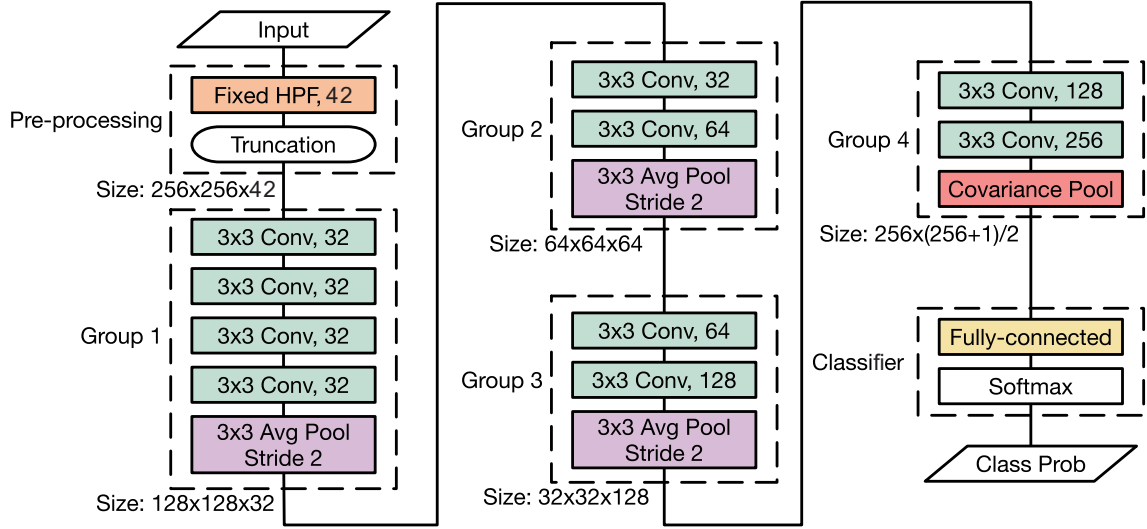}
    \caption{The CNN based steganalytic model used in the proposed framework}
    \label{fig:CNN_framework}
\end{figure*}

\section{Experimental Results and Analysis}
\label{sec:Results}

In our experiments, 20,000 uncompressed images of size $512\times 512$  are from BOSSBase-v1.01 \cite{bas2011break} and 
BOWS2 \cite{BOWS2}.   We firstly resize the images into $256\times 256$ using ``imresize'' in Matlab with default settings, and then divide the resized images into two non-overlapping parts randomly. The first part includes 10,000 images, which are used to train a  steganography model described in Section \ref{sec:Proposed}.   In this part,  we use 8,000 images for training, and the rest 2,000 images for the validation. The second part includes 10,000  images, which are used to generate JPEG stego images based on  the trained model.    For a given quality factor,  therefore, we finally obtain 10,000 cover-stego image pairs.  To evaluate steganography security,  we further divide the resulting 10,000 cover-stego image pairs into two non-overlapping parts.  The first part includes 5,000 image pairs, which are used to train a steganalytic classifier using the steganalytic model in Fig. \ref{fig:CNN_framework} and three modern steganalytic models (i.e. GFR \cite{GFR}, SCA-GFR \cite{SCA-GFR}, and SRNet \cite{SRNet}) separately.  The rest 5,000 image pairs are used for testing.  

In the following, we first describe the settings and parameters of the proposed steganography framework in our experiments, and then show the comparative results evaluated on different steganalytic classifers. Finally, we will provide some experimental analysis about the proposed steganography. 

\subsection{Settings of the Proposed Framework}
\label{subsec:Settings}

\subsubsection{\textbf{About Steganalytic Model}}
In the proposed  framework, we need to train a series of  CNN based steganalytic models.  To obtain a better tradeoff between the effectiveness and time complexity, we employ a modified version of our previous work \cite{Covariance2019} for image  steganalysis in spatial domain instead of the current best steganalyzer in JPEG domain, i.e. SRNet \cite{SRNet}.  As illustrated in Fig. \ref{fig:CNN_framework},  the CNN based architecture includes three modules,  that is,  the pre-processing, 4 layer groups and a classifier. The pre-processing module consists of 42 high-pass filters used in SRM and a truncation function.  The 4 layer groups have a similar structure, that is, several convolutional layers followed by a pooling layer.  Group 1 contains 4 convolutional layers while group 2-4 contain 2. The convolutional layers are of size $3\times  3$ and followed by a batch normalization layer and a ReLU activation function. Group 1-3 use an average pooling of size $3\times  3$  with the stride 2, while group 4 employs a global covariance pooling. The classifier consists of a fully-connected layer and a softmax function.  

Based on our experiments, we find that such a steganalytic model can achieve similar results to the SRNet \cite{SRNet}, while the time complexity is significantly decreased.

\subsubsection{\textbf{About Parameter Settings}}
There are several important parameters in the proposed steganography scheme,  including the iteration number $T$ of adversarial attacks,  the percentage $p$ for the selected gradients with larger amplitude,   and the adversarial intensity  $\alpha$ for adjusting  embedding cost.   Typically,  it is  time-consuming to obtain the best parameter combination of  $(T, p, \alpha )$ with brute-force searching.  In our experiments, we employ a greedy way.  
First of all,  we limit $T$ is no larger than 16,  since the detection accuracy will not drop significantly when $T$ is larger than 12  based on our experiments.
To obtain a better $(p,\alpha)$,   we first fix $\alpha=2.5$,  and select the best parameter $p$ in the set of \{0.1, 0.2, 0.3, 0.4, 0.5, 0.6, 1\} according to security performances in the first 8 rounds of iteration.  When the best $p$ is selected, we further evaluate the performances with different $\alpha$ in the 16 rounds of iteration. Here, we limit  $\alpha$ in the set of \{1.5, 2.0, 2.5, 3.0\} .  

\begin{figure*} \centering
\subfigure[Parameter $p$ selection, QF = 95] {\includegraphics[width=3.5in]{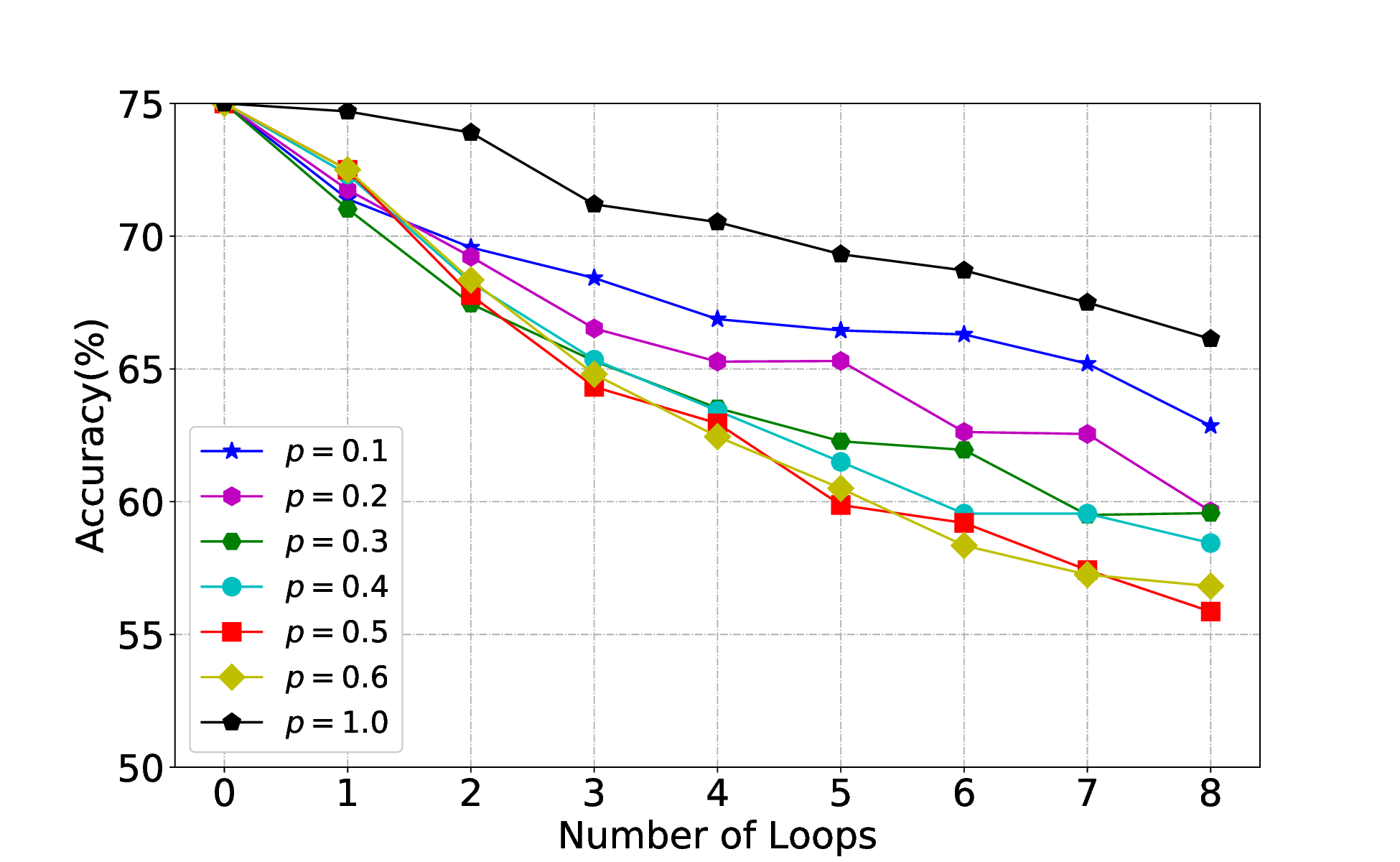}}
\subfigure[Parameter $\alpha$ selection, QF= 95] {\includegraphics[width=3.5in]{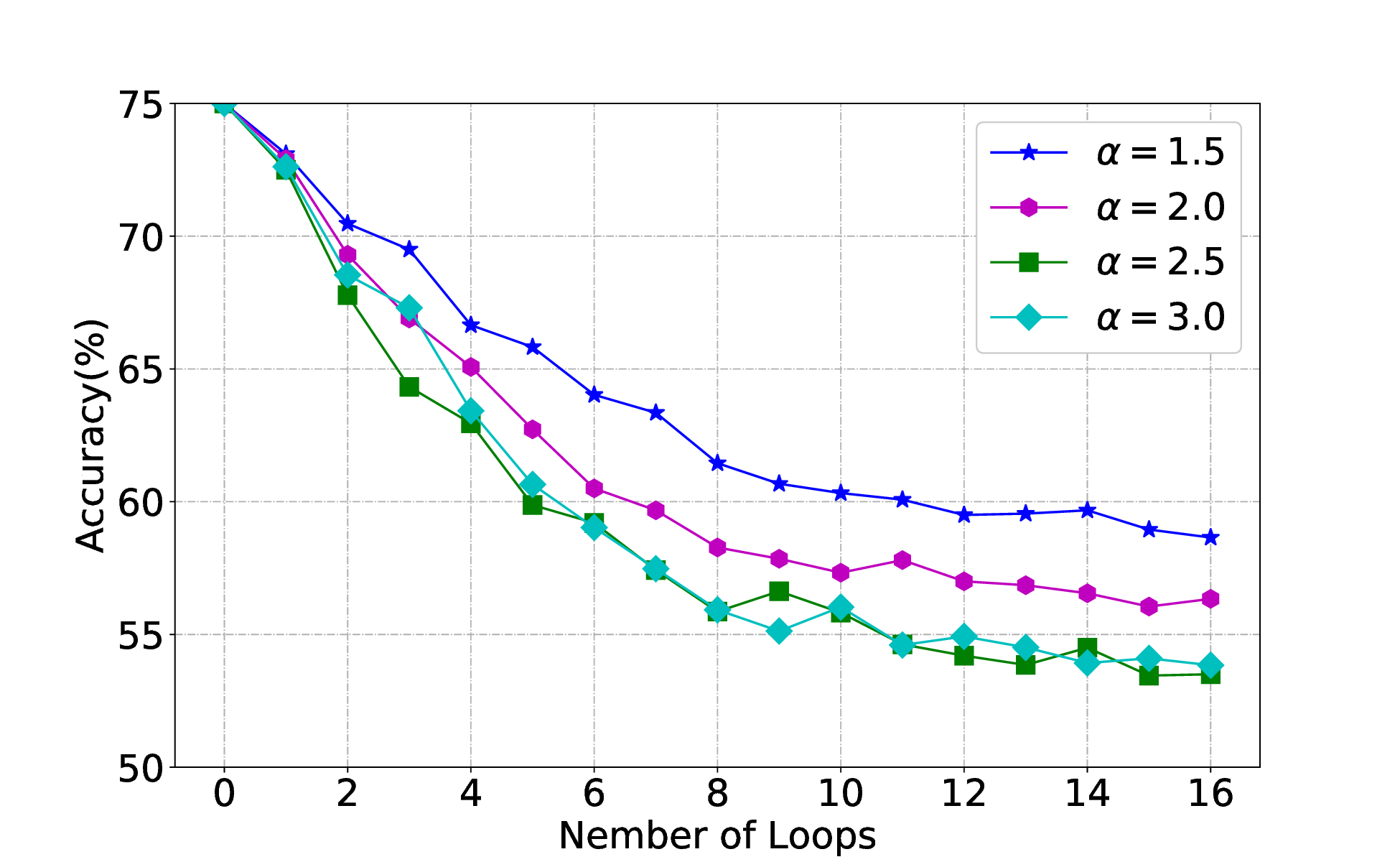}}
\\
\subfigure[Parameter $p$ selection, QF = 75] {\includegraphics[width=3.5in]{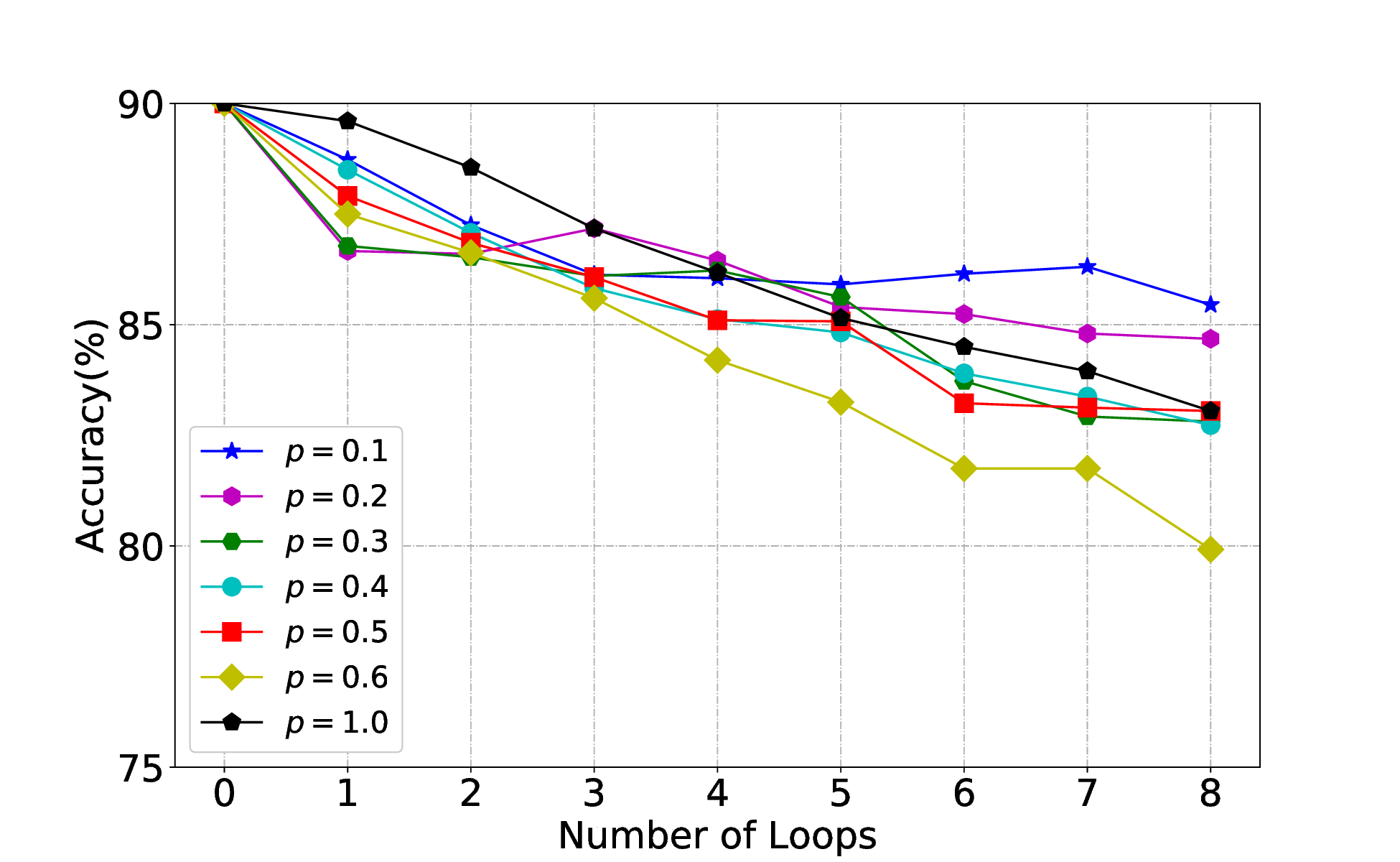}}
\subfigure[Parameter $\alpha$ selection, QF= 75] {\includegraphics[width=3.5in]{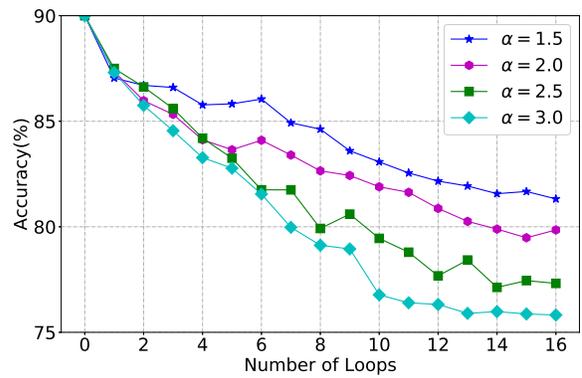}}
\caption{Detection accuracies evaluated on the validation set  (i.e. 2,000 cover-stego image pairs) with the increasing  number of loops. The value at the 0-th iteration is the detection accuracy on those images using original J-UNIWARD.}
\label{fig:parameters}
\end{figure*}

When payload is 0.40 bpnz,  some experimental results evaluated on the validation dataset are shown in Fig. \ref{fig:parameters}.  From Fig. \ref{fig:parameters},  we finally  determine that the proper parameter combination for  $QF = 95$  and $QF = 75$ is   $ \{T=16, p =0.50, \alpha=2.5\}$  and $\{T=16, p =0.60, \alpha=3.0\}$   separately. 

\subsection{Security Performances} 
\label{subsec:Security}

\begin{table*}[t!]
\caption{Detection accuracy (\%) evaluated on different steganalytic models. The embedding payload here is 0.40 bpnz, and  the value with an asterisk denotes the better result in the corresponding case.}
\label{table_1}
\centering
\begin{tabular}{c | c | c | c | c | c | c | c | c }
\hline \hline
\multirow{2}*{Quality Factor} & \multicolumn{2}{c|}{GFR} & \multicolumn{2}{c|}{SCA-GFR} & \multicolumn{2}{c|}{SRNet} & \multicolumn{2}{c}{Our Network in Fig.\ref{fig:CNN_framework}} \\ 
\cline{2-9}  
 & J-UNIWARD & Our Method &  J-UNIWARD &  Our Method  &  J-UNIWARD &  Our Method  &  J-UNIWARD &  Our Method  \\
\hline
95 & 63.51  & \textbf{53.04 *}  & 64.25  & \textbf{53.12 *} &  82.30 & \textbf{69.78 *} & 75.00  & \textbf{53.50 *}\\
\hline 
75 & 80.23  & \textbf{66.13 *} & 81.13  & \textbf{68.81 *}  & 91.50  & \textbf{85.33 *}  & 90.10  & \textbf{75.82 *} \\ 
\hline \hline
\end{tabular}
\label{tab:results}
\end{table*}

\begin{figure*}[t!]
    \centering
     \includegraphics[width=3.5in]{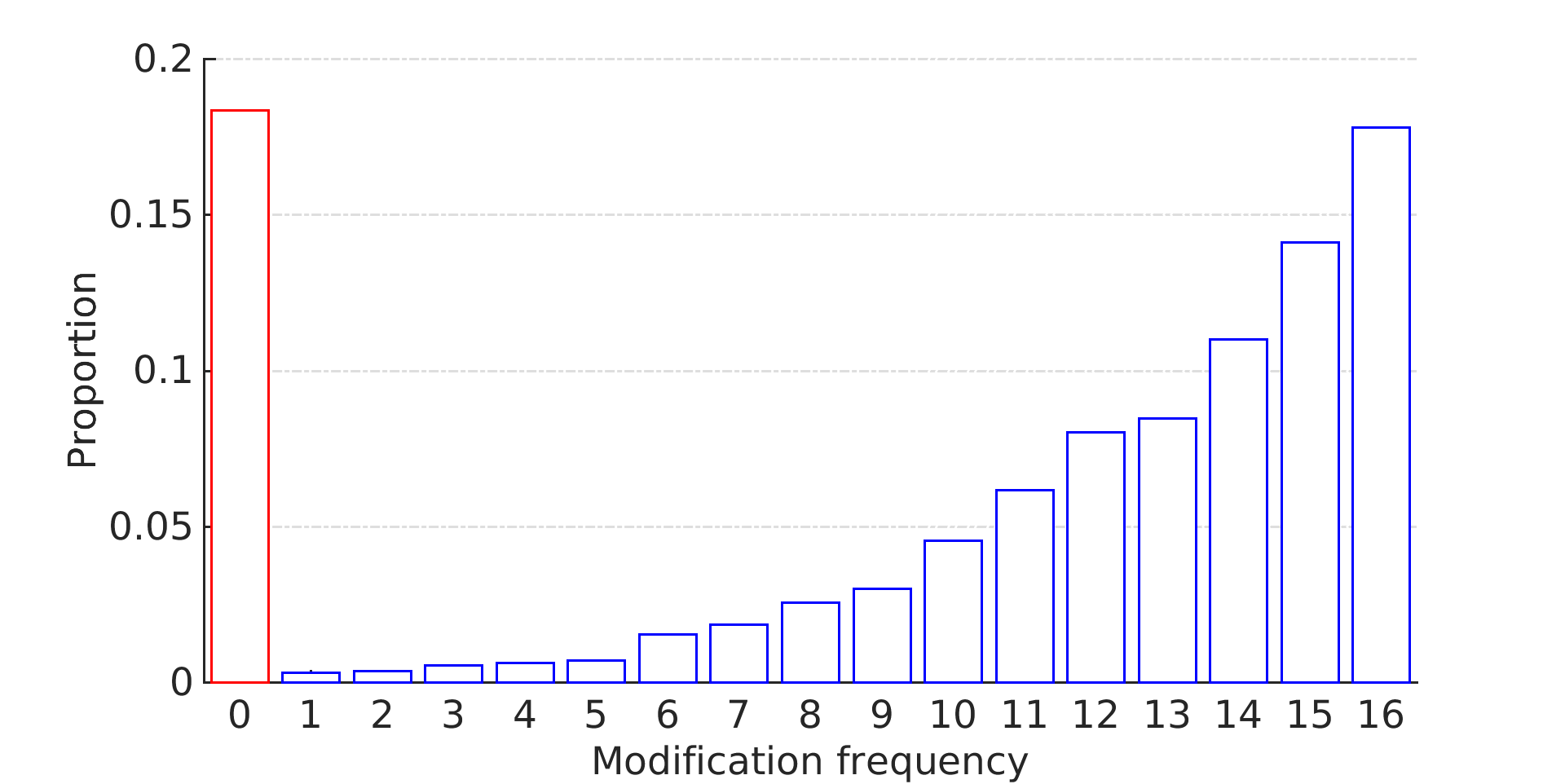}
      \includegraphics[width=3.5in]{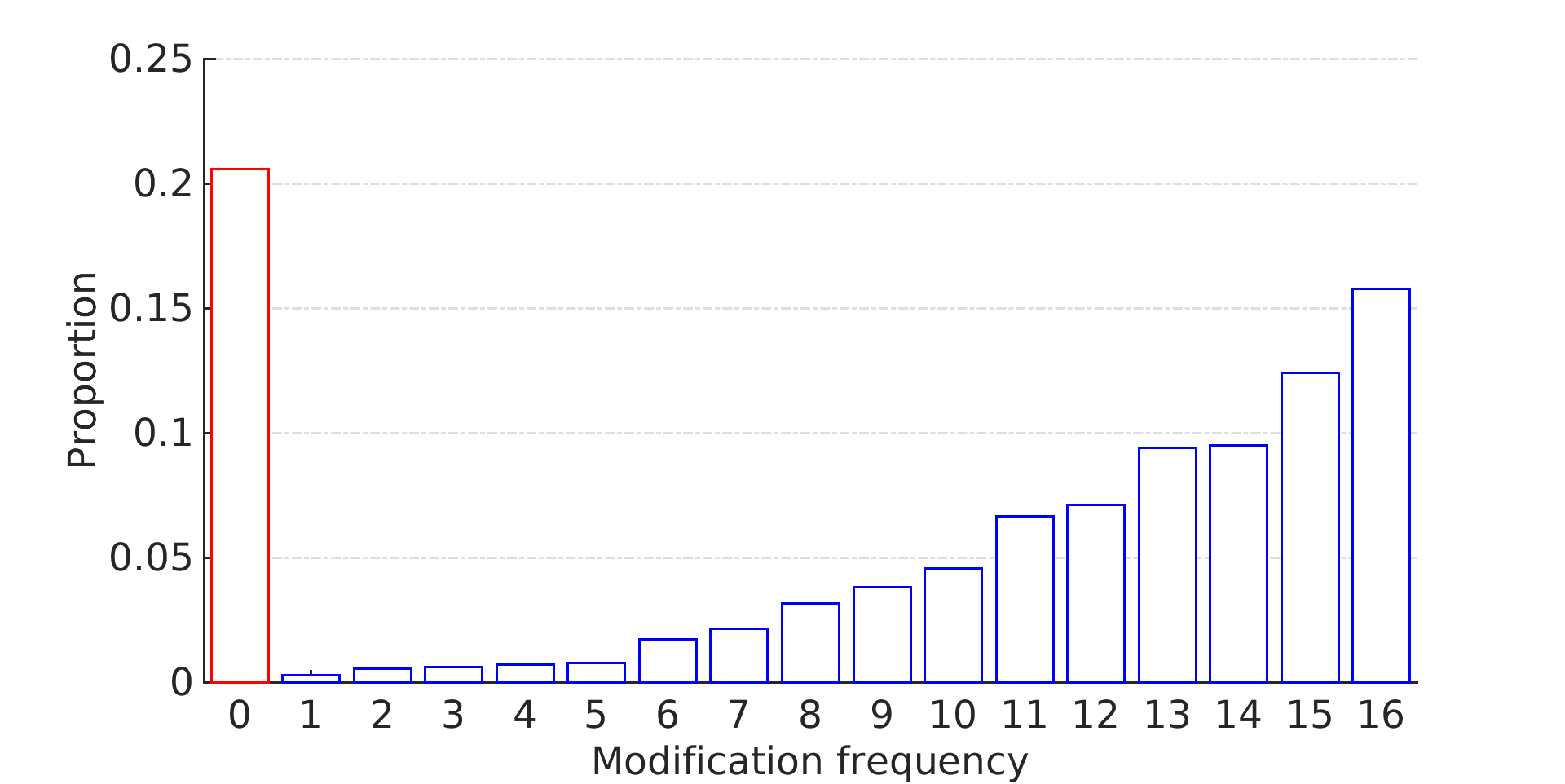}
    \caption{The histograms of the modification frequency of embedding costs on 10,000 test images with the quality factor 95 (left) and 75 (right). }
    \label{fig:Frequency}
\end{figure*}

Once the parameters are determined in previous section,  it is easily to generate stego image for any given cover as described in section \ref{subsec:Creating_Stegos}.   In this section, some  security performances evaluated on the test dataset will be given.  

The experimental results for payload 0.40 bpnz  are shown in Table \ref{tab:results}. From table \ref{tab:results},  we observe that the proposed framework can  improve the security performance of  J-UNIWARD on the four steganalytic models. For instance, when $QF=95$,  the detection accuracies with the current best steganalytic model SRNet are 82.30\% and 69.78\% for J-UNIWARD and our method separately, which means that there are over 12\% improvement in this case.  For the three other steganalytic models (i.e. GFR, SCA-GFR and the proposed network), we obtain over 10\%, 11\% and 21\% improvements  separately, which is a significant improvement in current JPEG steganography.  Similar results can be obtained when $QF=75$.  

\subsection{Analysis on the Modification of  Embedding Cost} 
\label{subsec:Analysis}
 
Since the proposed embedding cost can be regarded as a modified version of the J-UNIWARD, we would provide some experimental analysis on the modification of  embedding cost in this section. 

\subsubsection{\textbf{Modification Positions}} 
In each round of cost updating, we just consider those embedding costs whose gradients with larger amplitude because their changes would lead to faster change of the loss function based on the characteristic of gradients.  In this section, we will provide two  image examples to show the corresponding positions of the selected embedding costs with different parameter $p$.  

\begin{figure*}
\centering 
 \subfigure[Cover Example ]{\includegraphics[width=3.75cm]{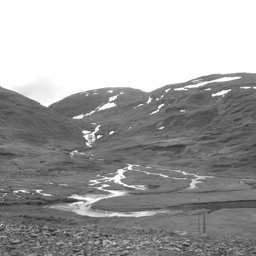}}  
  \subfigure[Top 10\%]{\includegraphics[width=3.75cm]{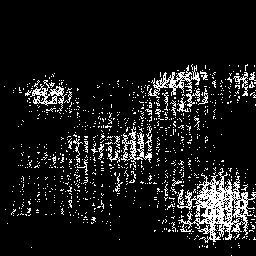}} 
 \subfigure[Top 30\%]{\includegraphics[width=3.75cm]{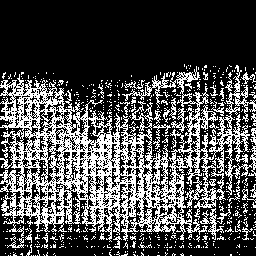}}
 \subfigure[Top 50\%]{\includegraphics[width=3.75cm]{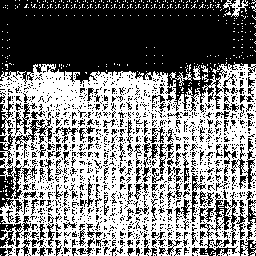}}  
 \\
  \subfigure[Cover Example ]{\includegraphics[width=3.75cm]{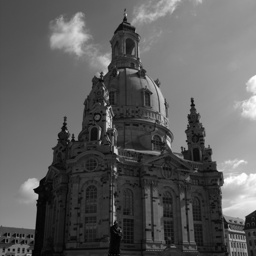}}  
  \subfigure[Top 10\%]{\includegraphics[width=3.75cm]{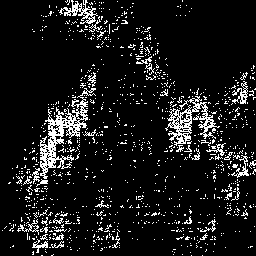}} 
 \subfigure[Top 30\%]{\includegraphics[width=3.75cm]{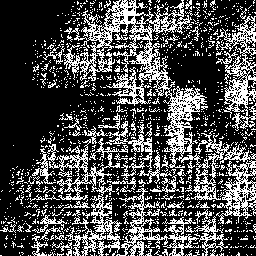}}
 \subfigure[Top 50\%]{\includegraphics[width=3.75cm]{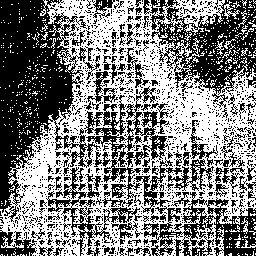}}  
\caption{The  positions of embedding costs to be updated  according to the selected gradients with  parameter $p$. The quality factor here is 95.}
  \label{fig:selected}
\end{figure*}

As illustrated in Fig.  \ref{fig:selected},  it is interesting to see  that  those regions with high complexity usually have larger gradient amplitude.  Similar results can also be observed from other image examples based on our experiments.  Therefore, the proposed method prefers to update those  embedding cost that located at the image regions with higher complexity,  and preserve smoother regions as they are.

\subsubsection{\textbf{Modification Frequency}}
In our experiments, the initial embedding costs (i.e. J-UNIWARD) will be updated 16 times using iterative adversarial examples.  
For each embedding cost,  thus, its frequency of being modified with the proposed method is ranging from 0 to 16. 

Fig. \ref{fig:Frequency}  shows the histograms of the modification frequency on all costs in 10,000 test images with the quality factors 95 and 75 separately.   From Fig. \ref{fig:Frequency}, we observe that for both quality factor,  around 20\% of the costs will not changed (see the red bar) at all.  While the proportion would  increase from around 0.30\% to  around 16.80\% with increasing the modification frequency form 1 to 16,  which means that most of the initial embedding costs have been repeatedly updated.

\begin{figure}[t!]
    \centering
     \includegraphics[width=2.75in]{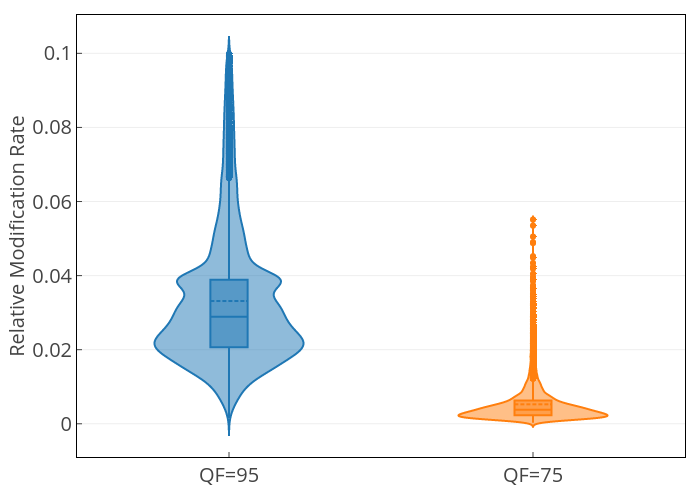}
    \caption{The relative modification rates on 10,000 test images}
    \label{fig:Modification}
\end{figure} 

\subsubsection{\textbf{Relative Modification Rate}}

For each embedding cost, the  absolute modification is ranging from 0 to $16\times \alpha$,  and around 80\% of the initial costs have been modified  with the proposed method from Fig. \ref{fig:Frequency}.   To measure the modification of an image quantitatively, we define the relative modification rate  based on the following formula:  

\begin{equation}
\label{eqa:6}
 \frac{\sum |\rho^{+}_c-\rho_{c,0}| + \sum |\rho^{-}_c-\rho_{c,0}|  }{2 \times \sum\rho_{c,0}}  
\end{equation} 

where $\rho_{c,0}$ and $\rho_c = (\rho^{+}_c,\rho^{-}_c) $ denote the embedding costs of J-UNIWARD and our method.   

Fig. \ref{fig:Modification} shows the violin-plot of the relative modification rates for  10,000 test images.  From Fig. \ref{fig:Modification}, we observe  that  although the proposed method would modify most embedding costs several times,  the relative modification rate is rather  lower.  On average, the relative modification rate is  3.31\% and 0.53\% for QF = 95 and QF = 75 separately, which is an interesting result that we can significantly improve the security performance of J-UNIWARD via modifying their embedding costs slightly.

\section{Conclusion}
\label{sec:Conclusion}
In this paper, we propose a novel  framework to enhance  JPEG steganography using iterative adversarial examples.   The main contributions of this paper are as follows.  
\begin{itemize}

\item  Unlike  existing steganography works based on adversarial examples,  we introduce a new framework to update JPEG embedding costs according to selected  gradients in an iterative manner.   The proposed framework is universal, which is also expected to be effective in other steganography, such as image in spatial domain. 

\item The proposed framework  can significantly enhance the security performances of  the existing JPEG steganography,  especially for the targeted CNN-based steganalyer employed in the framework.  

\item We analyze two important factors (i.e. $p$ and $\alpha$) that would significantly affect the proposed steganography framework,  and provide some statistical characteristics of the proposed embedding costs.

\end{itemize}
 
There are many issues worth further studying.   For instance, the proposed framework is still rely on an existing cost, i.e. J-UNIWARD.  The convergence property and the convergence efficiency on other embedding costs such as  flat or  random cost  will be included.  Besides,  just one steganalytic model is used in the proposed framework.  Some new schemes combined with  different steganalytic networks and/or stego post-processing \cite{Post-stego} would be considered to further improve the speed and security performance of the proposed framework.


\bibliographystyle{IEEEtran}
 \bibliography{cited}

\begin{thebibliography}{10}
\providecommand{\url}[1]{#1}
\csname url@samestyle\endcsname
\providecommand{\newblock}{\relax}
\providecommand{\bibinfo}[2]{#2}
\providecommand{\BIBentrySTDinterwordspacing}{\spaceskip=0pt\relax}
\providecommand{\BIBentryALTinterwordstretchfactor}{4}
\providecommand{\BIBentryALTinterwordspacing}{\spaceskip=\fontdimen2\font plus
\BIBentryALTinterwordstretchfactor\fontdimen3\font minus
  \fontdimen4\font\relax}
\providecommand{\BIBforeignlanguage}[2]{{%
\expandafter\ifx\csname l@#1\endcsname\relax
\typeout{** WARNING: IEEEtran.bst: No hyphenation pattern has been}%
\typeout{** loaded for the language `#1'. Using the pattern for}%
\typeout{** the default language instead.}%
\else
\language=\csname l@#1\endcsname
\fi
#2}}
\providecommand{\BIBdecl}{\relax}
\BIBdecl

\bibitem{szegedy2013intriguing}
C.~Szegedy, W.~Zaremba, I.~Sutskever, J.~Bruna, D.~Erhan, I.~Goodfellow, and
  R.~Fergus, ``Intriguing properties of neural networks,''
  \emph{arXiv:1312.6199}, 2013.

\bibitem{holub2014universal}
V.~Holub, J.~Fridrich, and T.~Denemark, ``Universal distortion function for
  steganography in an arbitrary domain,'' \emph{EURASIP Journal on Information
  Security}, vol. 2014, no.~1, p.~1, 2014.

\bibitem{HILL}
B.~Li, M.~Wang, J.~Huang, and X.~Li, ``A new cost function for spatial image
  steganography,'' in \emph{2014 IEEE International Conference on Image
  Processing (ICIP)}.\hskip 1em plus 0.5em minus 0.4em\relax IEEE, 2014, pp.
  4206--4210.

\bibitem{UERD}
L.~Guo, J.~Ni, W.~Su, C.~Tang, and Y.-Q. Shi, ``Using statistical image model
  for jpeg steganography: uniform embedding revisited,'' \emph{IEEE
  Transactions on Information Forensics and Security}, vol.~10, no.~12, pp.
  2669--2680, 2015.

\bibitem{STC}
T.~Filler, J.~Judas, and J.~Fridrich, ``Minimizing additive distortion in
  steganography using syndrome-trellis codes,'' \emph{IEEE Transactions on
  Information Forensics and Security}, vol.~6, no.~3, pp. 920--935, 2011.

\bibitem{Covariance2019}
X.~Deng, B.~Chen, W.~Luo, and D.~Luo, ``Fast and effective global covariance
  pooling network for image,'' in \emph{Proceedings of the 7th ACM Workshop on
  Information Hiding and Multimedia Security}.\hskip 1em plus 0.5em minus
  0.4em\relax ACM, 2019.

\bibitem{SRNet}
M.~Boroumand, M.~Chen, and J.~Fridrich, ``Deep residual network for
  steganalysis of digital images,'' \emph{IEEE Transactions on Information
  Forensics and Security}, vol.~14, no.~5, pp. 1181--1193, 2018.

\bibitem{goodfellow2015explaining}
I.~Goodfellow, J.~Shlens, and C.~Szegedy, ``Explaining and harnessing
  adversarial examples,'' in \emph{International Conference on Learning
  Representations}, 2015.

\bibitem{ma2018weakening}
S.~Ma, Q.~Guan, X.~Zhao, and Y.~Liu, ``Weakening the detecting capability of
  cnn-based steganalysis,'' \emph{arXiv preprint arXiv:1803.10889}, 2018.

\bibitem{Adversarial_IFS2019}
W.~Tang, B.~Li, S.~Tan, M.~Barni, and J.~Huang, ``{CNN}-based adversarial
  embedding for image steganography,'' \emph{IEEE Transactions on Information
  Forensics and Security}, 2019.

\bibitem{zhang_adversarial2018}
Y.~Zhang, W.~Zhang, K.~Chen, J.~Liu, Y.~Liu, and N.~Yu, ``Adversarial examples
  against deep neural network based steganalysis,'' in \emph{Proceedings of the
  6th ACM Workshop on Information Hiding and Multimedia Security}.\hskip 1em
  plus 0.5em minus 0.4em\relax ACM, 2018, pp. 67--72.

\bibitem{bas2011break}
P.~Bas, T.~Filler, and T.~Pevn{\`y}, ````break our steganographic system'': The
  ins and outs of organizing boss,'' in \emph{Springer International Workshop
  on Information Hiding}, 2011, pp. 59--70.

\bibitem{BOWS2}
P.~Bas and T.~Furon, ``Bows-2,'' \url{http://bows2.ec-lille.fr//}, 2007.

\bibitem{GFR}
X.~Song, F.~Liu, C.~Yang, X.~Luo, and Y.~Zhang, ``Steganalysis of adaptive jpeg
  steganography using 2d gabor filters,'' in \emph{Proceedings of the 3rd ACM
  workshop on information hiding and multimedia security}.\hskip 1em plus 0.5em
  minus 0.4em\relax ACM, 2015, pp. 15--23.

\bibitem{SCA-GFR}
T.~D. Denemark, M.~Boroumand, and J.~Fridrich, ``Steganalysis features for
  content-adaptive jpeg steganography,'' \emph{IEEE Transactions on Information
  Forensics and Security}, vol.~11, no.~8, pp. 1736--1746, 2016.

\bibitem{Post-stego}
B.~Chen, W.~Luo, and P.~Zheng, ``Enhancing steganography via stego
  post-processing by reducing image residual difference,'' in \emph{Proceedings
  of the ACM Workshop on Information Hiding and Multimedia Security}, to
  appear, 2019.

\end{thebibliography}
 
\end{document}